\documentclass[12pt]{iopart}
\usepackage{graphicx}
\usepackage{subfigure}
\usepackage{xcolor}

\def\e{\begin{equation}}
\def\f{\end{equation}}
\def\-#1{{\bf #1}}

\def\.{\cdot}

\begin{document}

\title[]{{Role of the normal polarization in the far-field subwavelength imaging by a dielectric microsphere or microcylinder}}

\author{R.~Heydarian$^1$, C.~R.~Simovski$^{1,2}$}

\address{$^1$ Department of Electronics and Nano-Engineering, Aalto University, P.O. Box 15500, FI-00076 Aalto, Finland}
\address{$^2$ Faculty of Physics and Engineering, ITMO University, 199034, Birzhevaya line 16, Saint-Petersbug, Russia}

\ead{reza.heydarian@aalto.fi}

\vspace{10pt}
\begin{indented}
\item January 2020
\end{indented}

\begin{abstract}
Role of the normal polarization in the far-field subwavelength imaging granted by a dielectric microsphere or microcylinder is discussed
and the hypotheses explaining this experimental fact are suggested. One of these hypotheses is confirmed by exact numerical simulations.
This mechanism of the magnifying superlens operation is based on the excitation of creeping waves at a curved dielectric interface by a
normally polarized dipole. The set of creeping waves after their ejection from the surface
creates an imaging beam which may mimic either a Bessel beam or a Mathieu beam depending on the microparticle radius.
This mechanism corresponds to the asymmetric coherent illumination.
\end{abstract}

%
%
%
%
\ioptwocol

\section{Introduction}

Nanoimaging of objects in real time -- beyond scanning the substantial areas with strongly submicron tips connected to cantilevers -- is a very important branch of nanophotonics. A lot of top-level studies has been done in this field recently and pioneering techniques were developed, such as stimulated emission depletion \cite{STED}, awarded by the Nobel prize in chemistry (2014). However, in spite of advantages of this method, there are applications, especially in the biomedicine, where fluorescent labels in the object area are prohibited (see e.g. in \cite{So}). Label-free optical nanoimaging still evokes a keen interest, and so-called superlenses  (see e.g. in \cite{PRB}) are still a subject of an intensive research. In the present work, we concentrate on a technique which seems to be the most affordable and straightforward type of superlens - dielectric spherical or cylindrical microlens.

In work \cite{Hong} it was experimentally revealed that a simple glass microsphere operates as a far-field magnifying superlens -- a device which creates a far-field magnified image of an object with its subwavelength details detectable by a conventional microscope. The imaged area is rather small (several square microns) and centered by the optical axis of the microscope passing through the microsphere center. Even few square microns is an area much larger than the object field of a scanning near-field optical microscope (SNOM). Therefore, this technique promises a much faster imaging of the whole substrate than the use of SNOM. A direct analogue of the spherical or cylindrical dielectric microparticle (MP) operating in the superlens regime is a metamaterial hyperlens \cite{Hyper1,Hyper2,Hyper3,Hyper4}. However, dielectric MPs are available on the market and are incomparably cheaper than the hyperlenses. Therefore, several scienrific groups have explored this field since 2011 (see e.g. in \cite{Astratov,Kassamakov,Yang,Zhou, Maslov}).  In \cite{Kassamakov} a direct lateral resolution on the level $\delta=\lambda/6$ was complemented by the interferometric resolution $\delta=\lambda/10$ in the normal direction. Also, in this work it was shown via simulations that two dipoles located at the surface of a glass  microsphere of the dimensionless radius $kR=60$ with the gap $\delta=\lambda/6$ between them can be resolved (in simulations) if and only if they are excited with nearly opposite phases. To have opposite phases is impossible for two small closely located scatterers illuminated by a plane wave. Moreover, the dielectric scatterers resolved in \cite{Kassamakov} with the gap $\delta=\lambda/6$ were illuminated by an incoherent light. Further, the resolution $\delta=\lambda/15$ was achieved using a glass MP for two plasmonic scatterers \cite{Astratov}.
The theory of these papers could not explain these experimental results.

However, in order to properly exploit a novel technique, one obviously needs to understand its physics. Initially, the authors of \cite{Hong} assumed that this imaging is related with the phenomenon of so-called photonic nanojet (PNJ) \cite{PNJ}. The PNJ maintains a slightly subwavelength ($(0.3-0.5)\lambda$) effective width along a path that extends more than $2\lambda$ behind the MP . In works \cite{Taflove,Itagi,Taflove1} it was shown that the PNJ is a non-resonant phenomenon and results from the constructive interference of cylindrical or spherical harmonics excited inside the MP by a plane wave. For the (relative to the ambient) refractive index of the MP $n=1.4-2$ a whatever MP radius from $R=\lambda$ to $R=20\lambda$ corresponds to a sufficient amount of spatial harmonics which experience the constructive interference at the rear extremity of the MP. The studies also have shown that in the near vicinity of the rear edge point the package of evanescent waves is excited that grants to the waist of the wave beam a high local intensity. Authors of \cite{Hong} assumed: since a plane wave excites these evanescent waves in a MP, the evanescent waves excited by a closely located subwavelength scatterer should reciprocally convert into propagating waves and form a PNJ.  However, further studies (see e.g. in \cite{Astratov,Kassamakov,Yang,Lecler,Zhou, Maslov,Astratov1} etc.) have shown that a scatterer located near a MP does not produce a PNJ behind it. Moreover, in work \cite{Astratov1} it was noticed that the explanation of the superlens functionality of a dielectric MP via the evanescent waves \cite{Yang,Lecler} is disputable because in presence of evanescent waves the reciprocity principle is not reducible to the inversion of the wave propagation. Really, in both focusing and emitting schemes, the evanescent waves decay in the same directions -- from the rear point of the sphere. Therefore, the evanescent waves responsible for the subwavelength width of the PNJ waist in the focusing scheme cannot be linked to those excited by the imaged object in the emitting scheme. To confirm this point in \cite{Astratov1} the exact simulations of the point-spread function for the structures from \cite{Hong,Kassamakov,Yang,Lecler} were done and the subwavelength imaging was absent, though the PNJ in the reciprocal case manifested the subwavelength waist. The only difference in these simulations from the experiments and theoretical speculations in the cited works was replacement of the 3D MP (sphere) by the 2D one (cylinder) that also implies the 2D dipole source (dipole line parallel to the cylinder axis). One may believe that the 3D geometry grants specific mechanisms of superlens operation compared to the 2D one. However, this belief does not disable the argument against the explanation of the MP superlens operation involving the PNJ and reciprocity. Moreover, the PNJ is formed by a dielectric MP in both 3D and 2D geometries with similar efficiency \cite{Taflove1}. If the PNJ model of a 2D MP superlens does not work (it was clearly proved in \cite{Astratov1}) why it will work for the 3D superlens?

In works \cite{Zhou,Maslov} resonant mechanisms of subwavelength imaging by a dielectric MP were analyzed. One was related to whispering gallery resonances, another -- to the Mie resonances. In both cases, the wave packages responsible for subwavelength hot spots inside the particle experience the leakage and partial (quite weak) conversion into propagating waves. Two other resonant mechanisms were recently reported in works \cite{Cang} and \cite{Wang}. However, all these resonant mechanisms do not explain why a dielectric microsphere operates as a far-field superlens in a broad frequency range. In work \cite{Sundaram} a broadband subwavelength resolution in the incoherent light was theoretically obtained for a glass MP. However it was as modest as $\delta=\lambda/4$ and demanded the use of an exotic microscope with a solid immersion lens as an objective. This microscope has the f-number smaller than unity. It obviously implies the reduction of the  diffraction-limited image size of a  point source compared to the finest size granted by a usual microscope $\delta\approx 0.5\lambda$ \cite{Born}. This size is the radius of the Airy circle in the image plane and it is equal to the finest possible resolution of a microscope \cite{Born}. For microscopes with small f-numbers the Airy circle radius (and finest resolution) $\delta=\lambda/4$ does not require additional imaging devices \cite{Papaliolios}. However, in all experiments with microspheres offering the subwavelength resolution the standard microscopes were used. Moreover, the resolution was noticeably finer than $\delta=\lambda/4$ predicted in \cite{Sundaram} as a limit value.

In the present paper, we suggest a hypothesis that the superlens operation of a dielectric MP is related to its capacity to create a diffraction-free wave beam. This property should be common for both 2D and 3D geometries. The imaging beam results from the emission of a dipole source (a small scatterer) which is polarized normally to the surface of a MP. There are at least two mechanisms which result in the formation of the diffraction-free beam by a dielectric MP. The first one is an incoherent mechanism and corresponds to the creation of a radially polarized Gaussian beam. The second one demands the illumination by a laser light and the nonzero phase shift between two dipoles in order to resolve them. This mechanism of subwavelength imaging results in either Bessel or Mathieu imaging beam. This hypothesis is confirmed by exact numerical simulations. In the end, we discuss how our results match the available literature data and predict the existence of one more mechanism responsible for the subwavelength imaging by a dielectric MP.

\begin{figure*}[h!]
\centering

\includegraphics[width=0.8\textwidth]{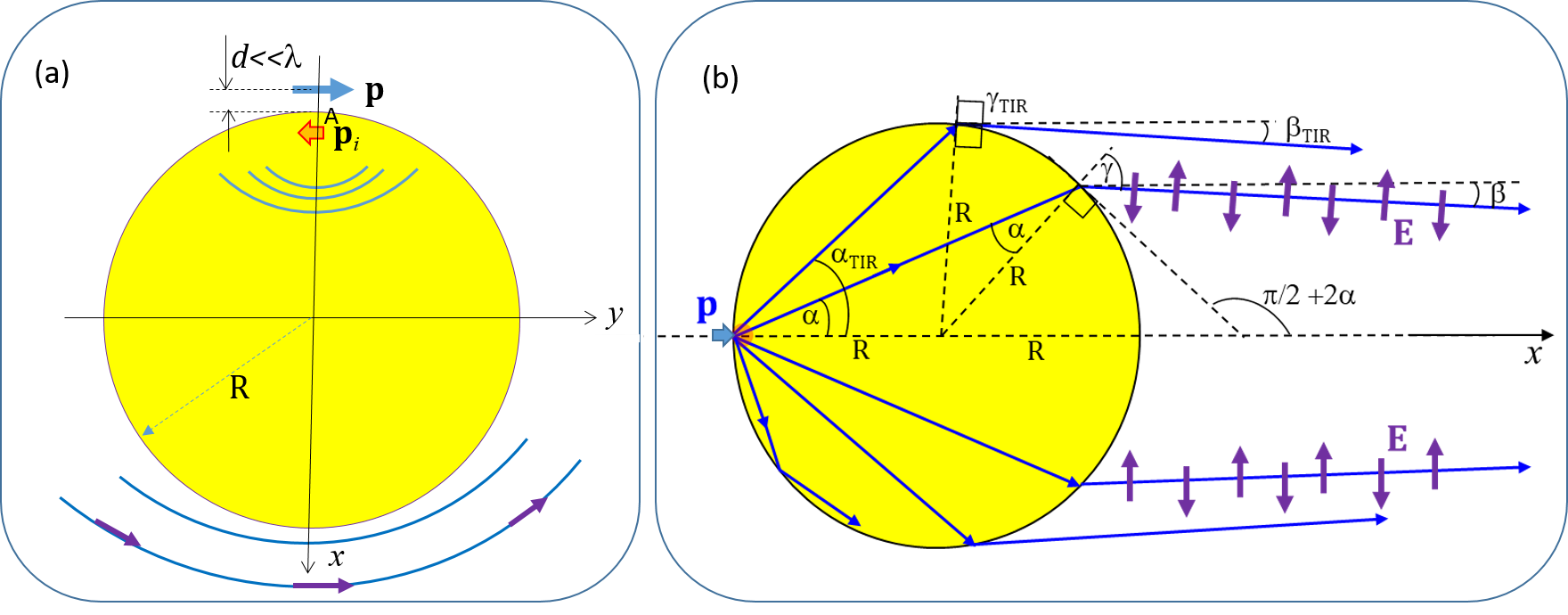}

\caption{(a) Radiation of a tangential dipole through a microsphere/microcylinder is weakly perturbed by a microsphere or a microcylinder. Thick blue lines show the wave fronts. (b) Radiation of a normal dipole through a microsphere or microcylinder within the framework of geometrical optics results in a radially polarized wave beam with low convergence. This convergence is zero ($\beta=\beta_{TIR}=0$) if $n=1.4$ (glass).}
\label{Pic1}
\end{figure*}

\section{Theory}

\subsection{Hypothesis of a radially polarized imaging beam}

When the unpolarized light impinges a dipole scatterer located near an MP ($kR\gg \pi$), the scatterer polarizes both tangentially to the particle surface and normally to it. To our knowledge, in all known works aiming to explain the superlens operation of a microsphere only a tangential dipole $\-p$ depicted in Fig.~\ref{Pic1}(a) was considered. It is difficult to expect the superlens operation in this case . For a subwavelength distance $d$ ($kd\ll \pi$) there is a near-field interaction resulting in the formation of the image dipole $\-p_i$ inside the MP at the distance $2d$ from point $A$. However, this dipole has the opposite phase with $\-p$. The radiation of this pair of dipoles transmits through the MP as it is described in work \cite{Astratov1} and represents a weakly directive wave beam experiencing the Abbe diffraction (the angular beam width grows with the distance).

For a normally polarized dipole, the situation depicted in Fig.~\ref{Pic1}(b) is different. The corresponding image dipole is in phase with the real one and the two dipole sources -- real and imaginary can be united into a dipole effectively located at the interface. In this case the imaging beam turns out to be almost diffraction-free.
Fig.~\ref{Pic1}(b) illustrates a simplistic model of the imaging beam -- that corresponding to the geometrical optics. Rays emitted by the total dipole with large angles of incidence to the rear interface of the sphere (refractive index $n$) experience total internal reflection (TIR). Only rays with $0<\alpha<\alpha_{TIR}$ transmit through the sphere. The tilt of the transmitted ray to the axis $x$ is equal $\beta=\gamma-2\alpha$, where $\gamma=\arcsin (n\sin\alpha)$. Calculating the derivative $\partial /\partial\beta$, it is easy to see that for $n>1.18$ there are no local maxima of $\beta$ for $0<\alpha<\alpha_{TIR}$ and the maximal tilt corresponds to the rays with
$\beta=\beta_{TIR}$. Moreover, for $n=1.4$ $\beta_{TIR}=0$ and all rays created by the dipole source are parallel to the axis $x$. It corresponds to the known fact that the focal point of a sphere with $n=1.4$ is located on its surface \cite{Lecler}. For a microsphere the radial polarization of the wave beam (illustrated in Fig.~\ref{Pic1}(b) by two sets of the vector $\-E$ on two symmetric rays) evidently follows from the problem geometry.

Gaussian beams with radial polarization are well known in the modern optics. They have the zero intensity at the optical axis and are formed as an eigenmode of an optical microfiber further transmitted into free space through the output cross section \cite{Tidwell}. In free space an ideal Gaussian beam with radial polarization
and uniform distribution over the azimuthal angle is not divergent (see e.g. in the overview \cite{radial}) i.e. does not experience the Abbe diffraction. If the scheme of an imaging beam depicted in Fig.~\ref{Pic1}(b) was fully adequate, two closely located dipoles would have created two radially polarized non-divergent wave beams. Due to the absence of the diffraction, these beams would interfere inside the MP where they intersect but would not form a single beam in which the information on the gap $\delta$ would be lost. On the contrary, they would propagate along the lines connecting the dipoles and the MP center. At a certain distance (that for usual wave beams would be the Fraunhofer diffraction zone) they would not intersect anymore and can be developed by a focusing lens forming two very distant images of two dipoles. A tightly focusing lens collects such a beam into a point where $\-E$ is polarized axially. In this scenario, the ultimate resolution granted by an MP is not diffraction-limited and the Airy circle has nothing to do with the resolution. The gap $\delta$ between two dipoles is in this scenario magnified by the factor $L/R$, where $L$ is the distance from the MP center to the objective plane. Of course, the point-wise is not achievable even in absence of the diffraction since it is restricted by optical noises \cite{Narimanov}. However, it is evident that it would be much finer than $\lambda/2$ for a standard microscope.

\subsection{Hypothesis of the imaging beam produced by creeping waves}

Of course, the geometric optical picture cannot be fully adequate for a microsphere whose radius $R$, though much larger than $\lambda$, is still comparable with it.
We can only aim to approach to this regime as closely as possible that we plan to do in future papers. In this work we study a different mechanism of the subwavelength resolution which demands the coherent illumination of the object.

\begin{figure*}[h!]
\centering
\includegraphics[width=0.8\textwidth]{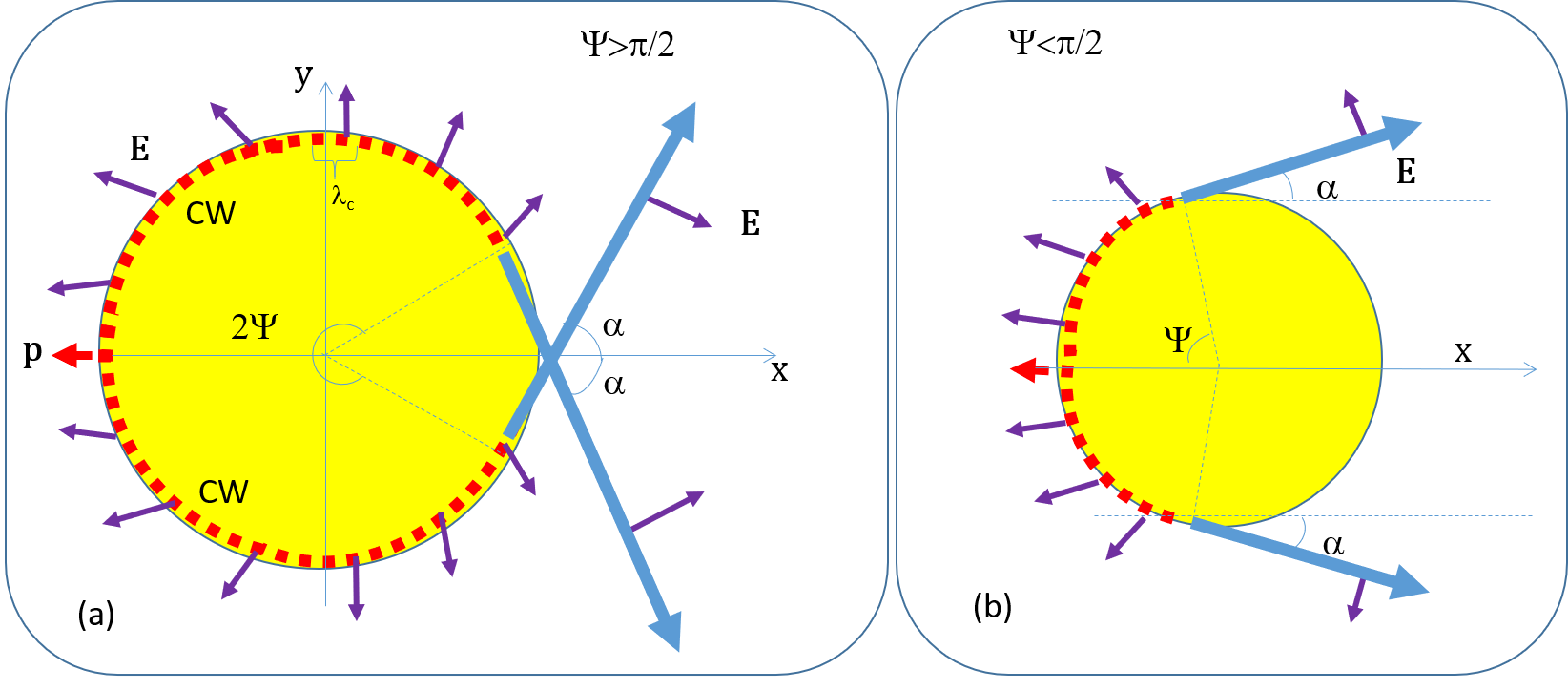}
\caption{An $m$-numbered creeping wave produced by a dipole after its leakage creates two partial beams ($\pm m$). For modest $kR$ the spectrum of CWs is rather broad whereas all $M$ CWs have the angular paths $\Psi$ larger than $\pi/2$. (b) The case of large $kR$ corresponds to all CWs having the paths $\Psi<\pi/2$.
Positive maxima of $\-E$ along the CW trajectory are shown as red squares.
}
\label{Pic2}
\end{figure*}

If a dipole scatterer is located closely to the MP ($kd\ll \pi$) the near-field coupling results in a very efficient excitation of creeping waves \cite{Felsen}.
The creeping waves (CWs) in the case of the normally oriented dipole are TM-polarized and propagate along the interface on its internal side.
Fig.~\ref{Pic2} can be referred to both 2D and 3D cases. For different sizes of our MP the wavenumbers of CWs can vary from $k_m\approx k$ to $k_m\approx k(n+1)/2$ \cite{Osipov}. For given $kR$ and $n$ a given source effectively excites a finite number $M$ of CWs \cite{Felsen,Osipov}.  If $kR\sim 10-20$ as in Fig.~\ref{Pic2}(a), all $k_m$ have real parts nearly equal to $k$, whereas the number $M$ of CWs is comparatively small and the angular paths $\Psi_m$ for all CWs from the birthplace to the ejection point exceeds $\pi/2$. If $kR\sim 50-100$ -- this case corresponds to Fig.~\ref{Pic2}(b) -- all $k_m$ have real parts close to $k(n+1)/2$, and the path of all CWs is rather short ($\Psi_m<\pi/2$), whereas $M$ is comparatively large. In \cite{Felsen} it is stressed, that in the case $kR\gg \pi$ the region where all CWs are ejected from the boundary is geometrically narrow (the angular width is as small as $5-10^{\circ}$).
As we can see in Fig.~\ref{Pic2}, in both cases one $m$-numbered CW forms two symmetrically tilted partial beams. One can be numbered $+m$ and another one can be numbered $-m$. Now, let us see what kind of image of our dipole is formed by such CWs.

\subsection{Imaging of one normally polarized dipole in creeping waves}

\begin{figure*}[h!]
\centering
\includegraphics[width=0.8\textwidth]{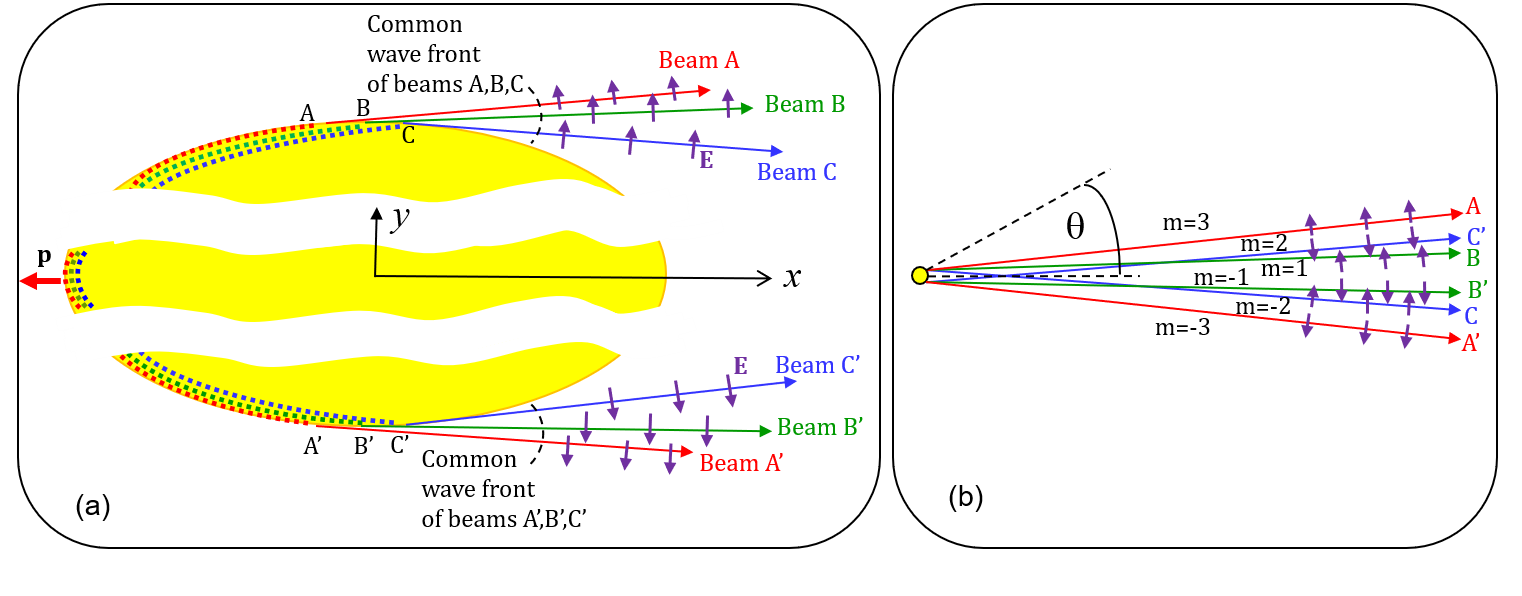}
\caption{Formation of a Bessel-like imaging beam consisting of $2M=6$ partial beams resulting from $M$ creeping waves. (a) Beams ejected from points A,B,C can be treated as three rays (nearly homocentric and emitted from the central point B of the ejection region ). The same refers to beams ejected from points A',B',C'. (b) In the Fraunhofer zone, the true phase of the electromagnetic field alternates across the imaging beam ($\pi$-jumps versus $m$).
Vectors $\bf E$ on every ray are shown in phase (with the interval $\lambda$).}
\label{Pic3}
\end{figure*}

As an example, consider the most interesting case when all $\Psi_m$ are close to $\pi/2$. In Fig.~\ref{Pic3}. Here we show the dipole creating three CWs and the regions where these CWs are ejected from the MP. Smaller path $\Psi_m$ corresponds to larger wave numbers $k_m$. Therefore, the beam ejected from point $A$ passing near the point $B$ has nearly the same phase as the beam ejected at $B$. The same refers to the beams ejected at points $B$ and $C$. Though the size of the ejection region (the distance $AC$) is of the order of $\lambda$ \cite{Felsen,Osipov}, three rays $A$, $B$ and $C$ are nearly homocentric as if they were
ejected from a subwavelength spatial region centered at $B$ {and have the common wave front as it is shown in Fig.~\ref{Pic3}(a)}. The same refers to the rays $A'$, $B'$ and $C'$.
In the 3D geometry, the wavefront corresponds to the radiation of a ring source of radius $R$.

Beams ejected from points B and B' have the smallest tilt and can be called partial beams of the first order ($m=\pm 1$), beams ejected from points C and C' are partial beams of the 2d order ($m=\pm 2$) and beams ejected from points A and A' are partial beams of the 3d order ($m=\pm 3$). Now let us take into account that the CW corresponding to the middle of their spectrum has maximal amplitude \cite{Osipov}, i.e. the electromagnetic field in the first-order partial beams (ejected from points $B$ and $B'$) is higher than that of the other partial beams. Assume that the electromagnetic field in the partial beams $C$ and $C'$ ($m=\pm 2$) is higher than that in the partial beams $A$ and $A'$ ($m=\pm 3$). Then our imaging beam qualitatively mimics the Bessel function of type $J_{\nu}$ of the argument $\xi \theta$, where $\theta$ is the tilt angle ($\nu$ and $\xi$ are parameters to be found). The true phase of the electromagnetic field in the partial beams jumps from $0$ to $\pi$ versus $m$ and it can be treated as the oscillation inherent to the Bessel function describing the electromagnetic field ($\-E$ and $\-H$) of our imaging beam. The first maxima of the Bessel function of $\theta$ are positive and correspond to the directions of the first-order beams $m=\pm 1$. In these partial beams, the electric field polarized along the polar vector  $\-\theta_0$ is adopted positive.
The directions of the beams $m=\pm 2$ correspond to the negative maxima of the Bessel function. In these partial beams, the electric field polarized along $\-\theta_0$ is adopted negative. Subtracting $(m-1)\pi$ from the true phase of each partial beam we may introduce its common effective phase and the common phase front. It is evident, that this phase front has a non-uniform curvature versus the polar angle $\theta$. Therefore, the imaging beam produced by a point dipole cannot be focused to a single subwavelength spot. In accordance to Fig.~\ref{Pic3}(b), our point dipole will be imaged as two parallel line sources in the 2D geometry and as a circle in the 3D geometry. In both cases the imaging beam reproduces the perimeter of the MP.

However, realistic Bessel beams used in modern optics though differ from an ideal Bessel beam exactly described by a Bessel function $J_{\nu}(\xi \theta)$, still grant the suppression of the Abbe diffraction by orders of magnitude \cite{Simon}. The negligibly small diffraction in the imaging beam created by the CWs though does not offer a subwavelength image in the incoherent light still allows, to our opinion, a subwavelength resolution. The hypothetic mechanism of this resolution will be presented in the next subsection.

To conclude this part, let us notice that the Bessel beam can be mimicked by our imaging beam only if there are CWs having both $\Psi>\pi/2$ and $\Psi<\pi/2$. In the 2D geometry it occurs when $kR=20-30$ and $n=1.7$. In this case, the distribution of the electromagnetic field across the imaging beam (simulated below) mimics the Bessel functions with the indices $\nu=0.5-1.5$ and an argument proportional to $\theta$. If $\Psi<\pi/2$ or $\Psi>\pi/2$ for all CWs, the true phases of all partial beams tilted upward in the geometry of Fig.~\ref{Pic3} will be positive and the true phases of all partial beams tilted downward will be negative on the common phase front of the imaging beam. In our numerical simulations, we observed $\Psi>\pi/2$ when $n=1.4,\, kR=10$ and $n=1.7,\, R=10-20$, whereas $\Psi<\pi/2$ corresponds to $kR=20-30$ and $n=1.4$.
In all these cases, the imaging beam turned out to be a Mathieu-like one. Realistic Mathieu light beams can be also considered as practically diffraction-free ones up to centimeter distances from their source \cite{Simon}.

\subsection{\label{dblcol}Imaging of a pair of normally polarized dipoles}

\begin{figure*}[h!]
    \centering

    \includegraphics[width=0.9\textwidth]{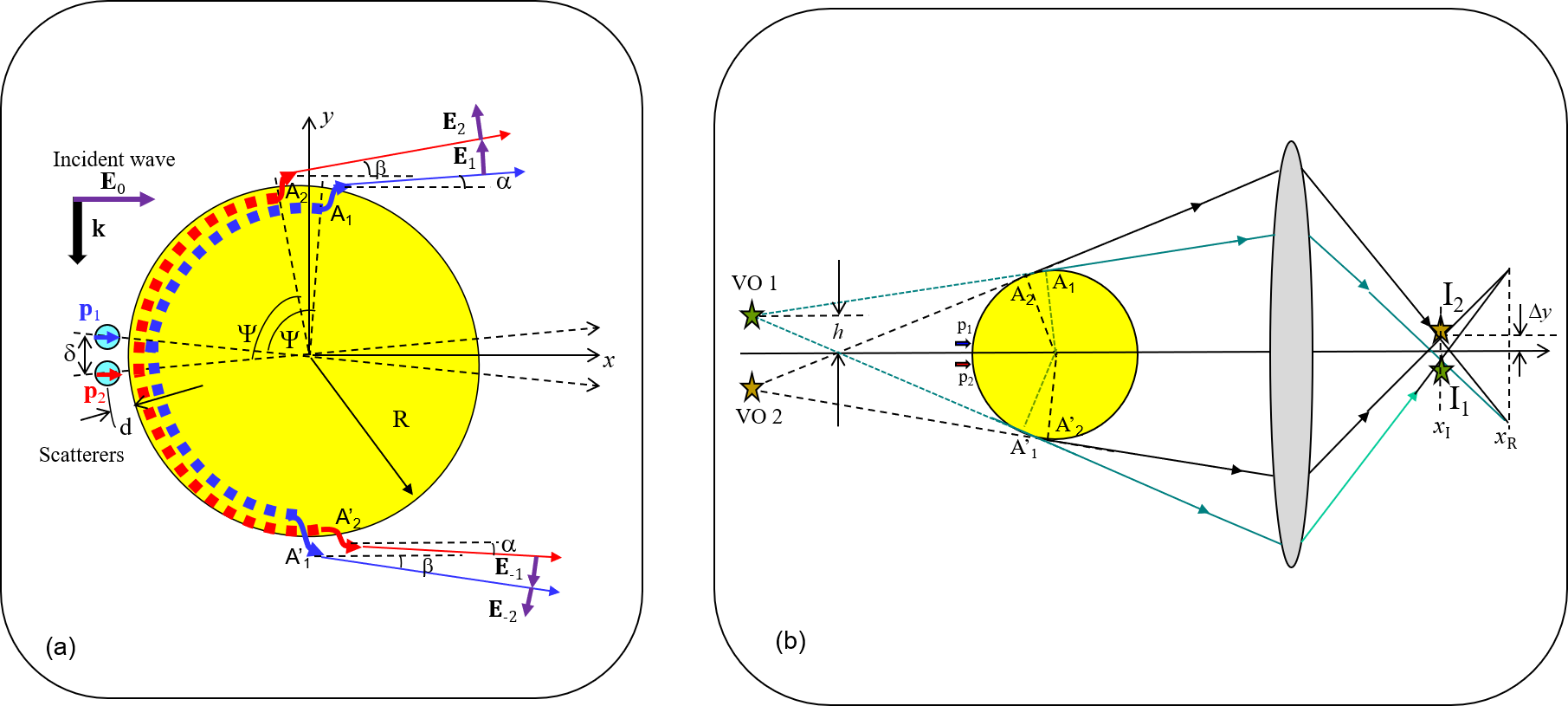}
    \caption{The phase shift between two point dipoles proportional to their separation results in the magnified image. (a) Two small scatterers 1 and 2{,} illuminated by a plane wave{,} acquire dipole moments $\-p_{1,2}$ are produce  CWs in the microparticle. Two CWs of the same order imply a birefringence of the imaging wave beam compared to that produced by one solid scatterer. Since $p_2=p_1e^{i\phi}$ two in-phase rays (the same color in our drawing) have different tilt angles. (b) Each pair of in-phase rays emitted from top and bottom edges of the microparticle with different tilt angles meet one another with the same phase at points $I_1$ and $I_2$ separated by a substantial gap $2\Delta y$. At both these points $M$ partial beams intersect being nearly focused. Therefore, points $I_1$ and $I_2$ are local maxima of light intensity.}

    \label{Pic4}
\end{figure*}

Consider a pair of dipoles $\-p_{1,2}$ separated by a subwavelength gap $\delta$ and located at a subwavelength distance $d$ from the MP.
If these dipoles are induced in two identical scatterers $1$ and $2$ by the incident wave polarized along $x$ as it is shown in Fig.~\ref{Pic4}, it is basically the same
as the polarization of $\-p_{1,2}$ normal to the surface of the MP, whereas the absolute value of $p_{1,2}$ of these dipole moments is the same.
Two CWs of the same order produced by these two dipoles results in two pairs of symmetrically tilted beams ejected from points $A_{1,2}$ and $A'_{1,2}$.
The angle between the beams ejected from points $A_1$ and $A_2$ ($A'_1$ and $A'_2$) equals $\delta/R$. This angle is much smaller than the angles between partial beams of different order $m$ and even smaller than the angular width of a partial beam corresponding to a given $m$. Therefore, if the phase shift $\phi$ is zero between dipole moments $p_1$ and $p_2$ (e.g. scatterers $1$ and $2$ are excited by a non-coherent light) we have the same fields $E_2=E_1$ on the effective phase fronts of the partial beams emitted from points $A_1(A'_1)$ and $A_2(A'_2)$. Here we imply the same phase for the vectors oriented so that the true phase of $E_y$ and $H_z$ at two symmetric partial beams differs by $\pi$. In other words, for a Bessel-like imaging beam we mean the Bessel phase that takes into account the $\pi$-jump of the true phase of $E_y$ and $H_z$ between two adjacent partial beams. For a Mathieu-like imaging beams we mean the Mathieu phase that implies $\pi$ subtracted from the true phase of $E_y$ and $H_z$ of all partial beams with $m<0$.

For a symmetric dual source $p_1=p_2$, a slight birefringence of the imaging beam granted by the gap $\delta$ between two dipoles has no noticeable implications for the imaging. It only results in a slight extension of the single dipole image corresponding to the total dipole $\-p_1+\-p_2$.
However, a coherent illumination illustrated by Fig.~\ref{Pic4}(a) implies $p_2=p_1\exp({i\phi})$, where $\phi=k\delta$. This situation is drastically different because we have $E_2=E_1\exp({i\phi})$ for two partial beams ejected from the top and  $E_{-1}=E_{-2}\exp({i\phi})$ for two partial beams ejected from the bottom. Two in-phase partial beams (the same color in our drawing) have different tilt angles.
The practical absence of the Abbe diffraction means that these partial beams though interfere do not mix up and form an anti-symmetric interference pattern in the top and bottom parts of the imaging beam.
For all partial beams emitted from the region $A_1A_2$ and for those emitted from the region $A'_1A'_2$ the phase distribution is anti-symmetric. The phase difference for a given tilt $\theta$ is equal $\phi$.

In Fig.~\ref{Pic4}(b) two pairs of in-phase partial beams (main maxima of the imaging beam of the Bessel or Mathieu type) meet one another in phase at points $I_1$ and $I_2$, which are, therefore, local maxima of intensity.
The coordinate $x_I$ of these points is close to the coordinate of the plane where the top and bottom parts of the imaging beam converge and the aforementioned image of the total dipole is formed. Therefore, around points $I_1$ and $I_2$ all partial beams are though not yet focused, but sufficiently converged so that their intensity at points $I_1$ and $I_2$ would be sufficient for imaging. Thus, we obtain two rather weak but distinguished images centered at points $I_1$ and $I_2$ which are distanced from one another by the macroscopic gap $2\Delta y$.
This gap is a magnified distance between two virtual objects $VO_1$ and $VO_2$ and equals to the product of the gap $2h$ between these points by the standard lens magnification factor $\Gamma$. $VO_1$ and $VO_2$ are effective phase centers from which the pairs of the in-phase beams are seemingly emitted.
Since $A_1A_2=\delta$, it is easy to see that $h=\delta/2\sin\Psi$ where $\Psi$ is the angular path of the CW from point $\-p_i$ to point $A_i$ ($A'_i$), $i=1,2$.
Since for given $kR$ paths $\Psi$ of different CWs differ weakly, in our estimation we may admit that the points $VO_{1,2}$ correspond to the mean angle $\Psi$ of the corresponding spectrum of CWs. Then we may write the result of our model in the form
\begin{equation}
\centering
\Delta y=\Gamma{\delta\over 2\sin\Psi}.
\label{h}
\end{equation}

Since $\sin\Psi$ is not very small the magnification of the dual dipole source is of the same order of magnitude as the standard magnification granted by the lens. The same refers to the image seen in a microscope.
In the 2D case, points $I_1$ and $I_2$ on the plane $xy$ mean the central lines of the strips of enhanced intensity. The local maximum of intensity at these points is granted by the constructive interference of any partial beam $+m$ with a beam $-m$ created by the same source and having therefore a different tilt. In the 3D case, the points $I_{1,2}$  are namely points of the maximal intensity and not traces of a ring. The ring imaging the total dipole is located in a different plane $x_R\ne x_I$ (distanced by several $\lambda$).
The gap between the dipoles $1$ and $2$ is located in the plane $xy$, the phase shift between symmetrically tilted rays holds namely in this plane and the corresponding maxima of intensity are formed only in this plane.
Thus, a conformal magnified image of two dipole scatterers separated by a subwavelength gap $\delta$ arises in both 2D and 3D cases.

Here, it is worth noticing that the virtual objects in Fig.~\ref{Pic4}(b) arise if and only if the dipoles $\-p_1$ and $\-p_2$ are out of phase. If the beams ejected from points $A_1$ and $A_2$ are in phase, the top and bottom parts of the imaging beam are homocentric and we will see only the image of the total dipole in the plane $x=x_R$. Moreover, these objects are located as it is shown in the drawing only in the case $\phi=k\delta$.

It is important that the areas of enhanced intensity centered by points $I_{1,2}$ are not images of separate dipoles $\-p_1$ and $\-p_2$. What is shown in Fig.~\ref{Pic4}(b) is a consolidate image of an asymmetric (phase-shifted) dual source. However, in accordance to formula (\ref{h}) it is a conformal image. It keeps conformal if the incidence of the illuminating wave is not grazing and the phase difference $\phi$ between the rays ejected from points $A_1$ and $A_2$ is equal $\phi=k\delta \cos\Phi$, where $\Phi$
is the incidence angle counted from the $y$-axis. In this case, the right-hand side in (\ref{h}) should be multiplied by $\cos\Phi$. When $\cos\Phi$ decreases maxima of intensity at points $I_{1,2}$ become weaker and for a certain $\Phi$ overlap. The minimal phase shift still granting the resolution as well as the minimal gap $\delta$
cannot be found from these qualitative speculations. They may be retrieved from exact simulations or found experimentally. However, this is not a subject of the present paper.
The main message of this theoretical part is the hypothesis of a key role of the object polarization in the direction orthogonal to the surface of the MP. We have assumed here that this polarization is responsible for the diffraction-free imaging wave beam created by a dielectric MP.

\section{Calculations: Results and Discussions}

In this section we report only the simulations of a 2D structure because for $kR\gg 10$ no one available simulator offers a reliable solution of the 3D problem (at least, in a reasonable computation time). Meanwhile COMSOL Multiphysics provides a rapid solver of 2D problems. It allowed us to obtain the color movies of the wave beams, color maps of their intensities and vector maps. Using COMSOL we are capable
to study the evolution of the wave beams up to hundreds $\lambda$. On the first stage, we checked the accuracy of the COMSOL solver reproducing the results obtained for a tangentially oriented dipole source in work \cite{Astratov1}.
We obtained the results depicted in Figs. 7 and 8 of \cite{Astratov1} for a glass cylinder with radiuses $kR=10, 20, 20.382, 30$ for a dipole located at the distance $d=1/k$. The case of whispering gallery resonance when $kR=20.382$ corresponds to the maximal image size.
We have further analyzed the diffraction spreading of the transmitted beam in the Fraunhofer zone of the MP and can confirm what is claimed in \cite{Astratov1} about the properties of a point-spread function.
We have seen no CWs excited by a tangential dipole in both glass MP ($n=1.4$) and in MP of transparent resin ($n=1.7$).

\subsection{Simulations for one normally polarized dipole line}

\begin{figure*}[h!]
\centering
\includegraphics[width=0.95\textwidth]{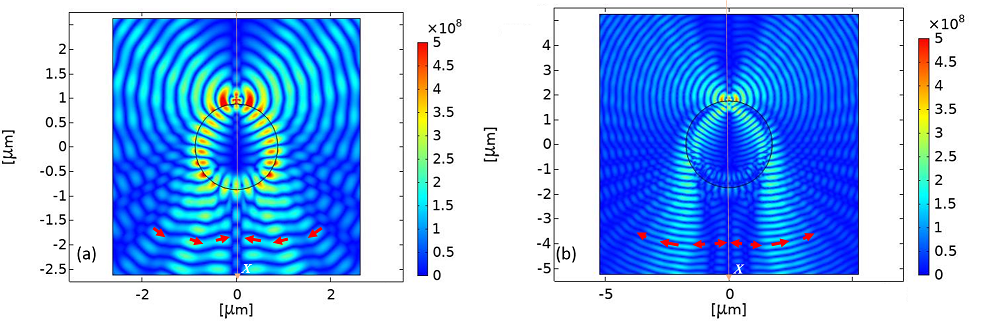}
\caption{Single radially polarized source -- wave pictures of the Fresnel zone for a glass microparticle with $kR=10$ (a) and $kR=20$ (b).
Red arrows show $\-E$ in some symmetrically located points of the wave picture. }
\label{Pic5}
\end{figure*}

We have performed extended numerical simulations of the structure with a single normally polarized dipole line source varying the radius of the MP in the range $kR=10-30$ for two values of the refractive index $n=1.4$ and $n=1.7$. In our simulations, the distance $d$ from the sources to the cylinder is equal $1/2k$. For the dual source the gap between the dipole lines was equal $\delta=2d$. The phase shift in the dual source in our simulations was varying from $0$ to $k\delta(n+1)/2$ (the results for $\phi=k\delta$ are most important).

In Fig.~\ref{Pic5} we present two instantaneous pictures of the wave movie for the cases $kR=10$, $n=1.4$ and $kR=20$, $n=1.4$. The patterns of dominating CWs are clearly seen. In the case $kR=10$ all CWs eject from the bottom part of the particle i.e. $\Psi>\pi/2$ as it was expected. The vertical ($x$-) component of the electric field has the same true phase at two points symmetrically located with respect to the axis $x$. The intensity is maximal in the partial beams of lowest order $m=\pm 1$. This beam mimics a Mathieu function and our studies of its evolution in the Franhofer zone confirm it. In the case $kR=20$ some CWs have the angular path $\Psi>\pi/2$ and some CWs have $\Psi<\pi/2$ as it was expected. For this case we expected to obtain a Bessel-like imaging beam.
However, in this particular case the imaging beam does not mimic any Bessel function -- the intensity in the partial beams $m=\pm 1$ is lower than that in the partial beams $m=\pm 2$ for which the intensity is maximal over $m$.
The distribution of the magnetic field $H_z$ in this case mimics the derivative of the Bessel function over the index: $\partial J_{\nu}(\xi\theta)/\partial \nu$, where $\xi\approx 15$ $1/rad$ and $\nu=1.5$. Such beams, to our knowledge, have never been studied. In our simulations we have not found the features of the Abbe diffraction for these beams, like other cases when the imaging beams mimic the Mathieu and Bessel beams.

\begin{figure*}[h!]
\centering
\includegraphics[width=0.95\textwidth]{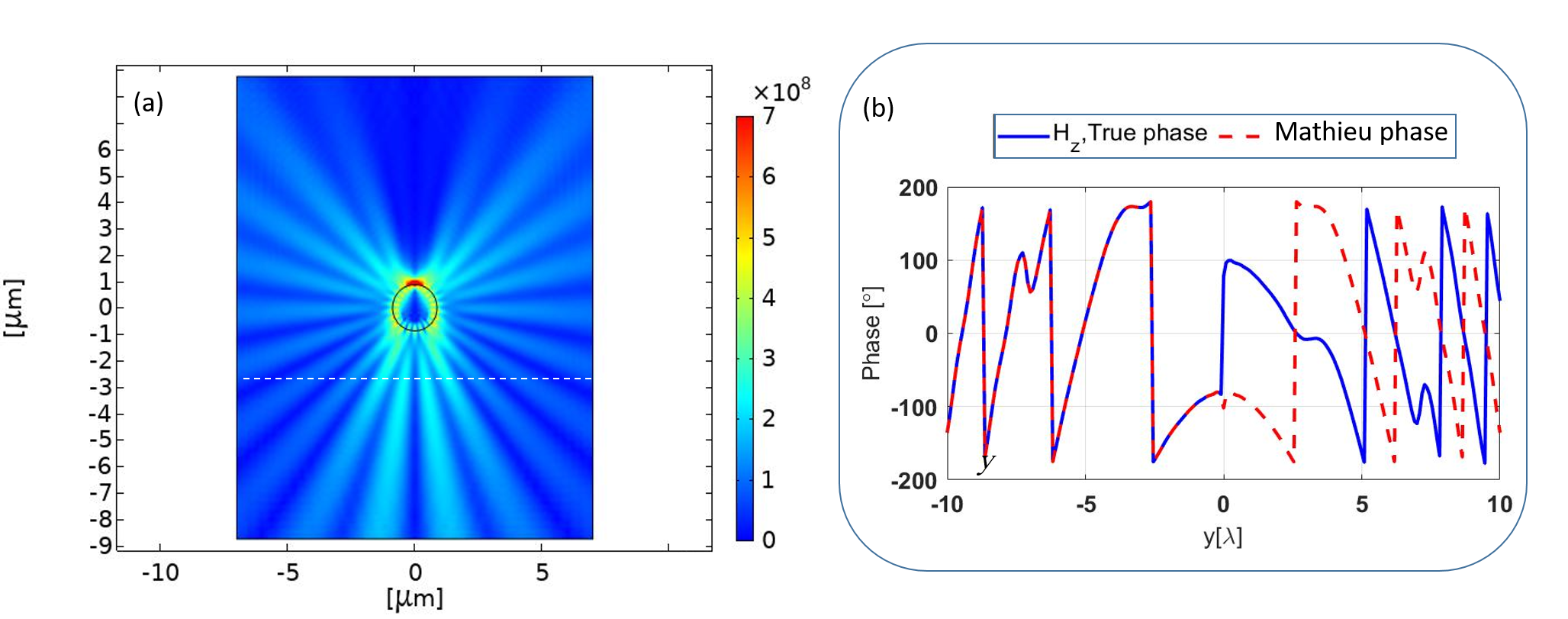}
\caption{Single source -- an intensity map in the optically large area (a) and a phase distribution across partial beams with $m=\pm 1,\pm 2,\pm 3$ (b).
White dashed line in (a) shows the argument of the phase distribution in (b). Glass microparticle with $kR=10$, $\lambda=550$ nm.
 }
\label{Pic6}
\end{figure*}

Fig.~\ref{Pic6}(a) shows that the effective angular width of all partial beams keeps the same at large distances from the MP corresponding to its Fraunhofer zone. There is no typical spread inherent to the Abbe diffraction. We have checked that the
Mathieu-like imaging beam (negative coordinates $x$ here correspond to the domain behind the cylinder) keeps practically diffraction-free up to $100-200\lambda$. We are sure that the diffraction is not similarly absent at larger distances up to macroscopic ones. In Fig.~\ref{Pic6}(b) we have shown the phase distribution of the magnetic field vector ($\-H=H\-z_0$) across four partial beams $m=\pm 1,\pm 2$. The true phase jumps by $\pi$ at the symmetry axis and subtracting this jump we see the symmetric phase distribution that we called above the Mathieu phase. Partial beams $m=\pm 1,\pm 2$ occupy the region $y=[-7,7]\lambda$ and we see a small oscillation of the phase corresponding to the dark area between the partial beams $m=\pm 2$ and $m=\pm 3$. Notice that the Mathieu phase is not constant across a partial beam because the dashed line shown in
Fig.~\ref{Pic6}(a) does not coincide with the phase front. The numerical retrieval of the phase front is doable but difficult and not relevant. Jumps  of the Mathieu phase equal to $2\pi$ are introduced in Fig.~\ref{Pic6}(b) to make the plot more compact.
Both pictures Fig.~\ref{Pic6}(a) and (b) correspond to our theoretical expectations.

\begin{figure*}[h!]
\centering
\includegraphics[width=0.95\textwidth]{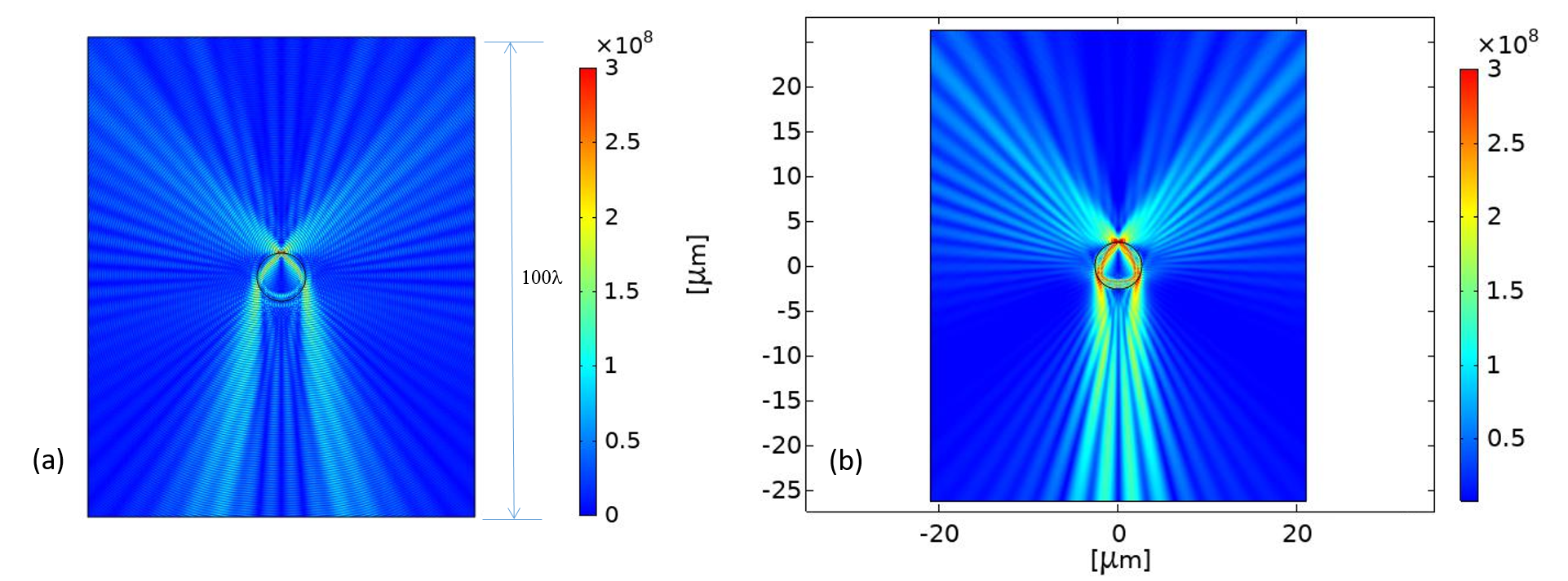}
\caption{Single source -- a wave picture in the large area for  $n=1.4$ (a) and
an intensity map in the same area for $n=1.7$ (b). Here $kR=30$ and $\lambda=550$ nm.
 }
\label{Pic7}
\end{figure*}

Similar conclusions refer to all our simulations for a single normally oriented dipole. Two more examples are presented in Fig.~\ref{Pic7}. Here in both pictures Fig.~\ref{Pic7}(a) and Fig.~\ref{Pic7}(b) we can see the mixture of two mechanisms of subwavelength imaging. In the wave picture
Fig.~\ref{Pic7}(a) we observe a weaker impact of CWs. Here a radially polarized imaging beam with low divergence as in Fig.~\ref{Pic2}(b) is presented together with the Mathieu-like beam resulting from CWs. The mechanism of CWs is more pronounced in Fig.~\ref{Pic7}(b).
The features of the diffraction were not found for both these cases until $100-200\lambda$.

\begin{figure*}[h!]
    \centering
    \includegraphics[width=0.95\textwidth]{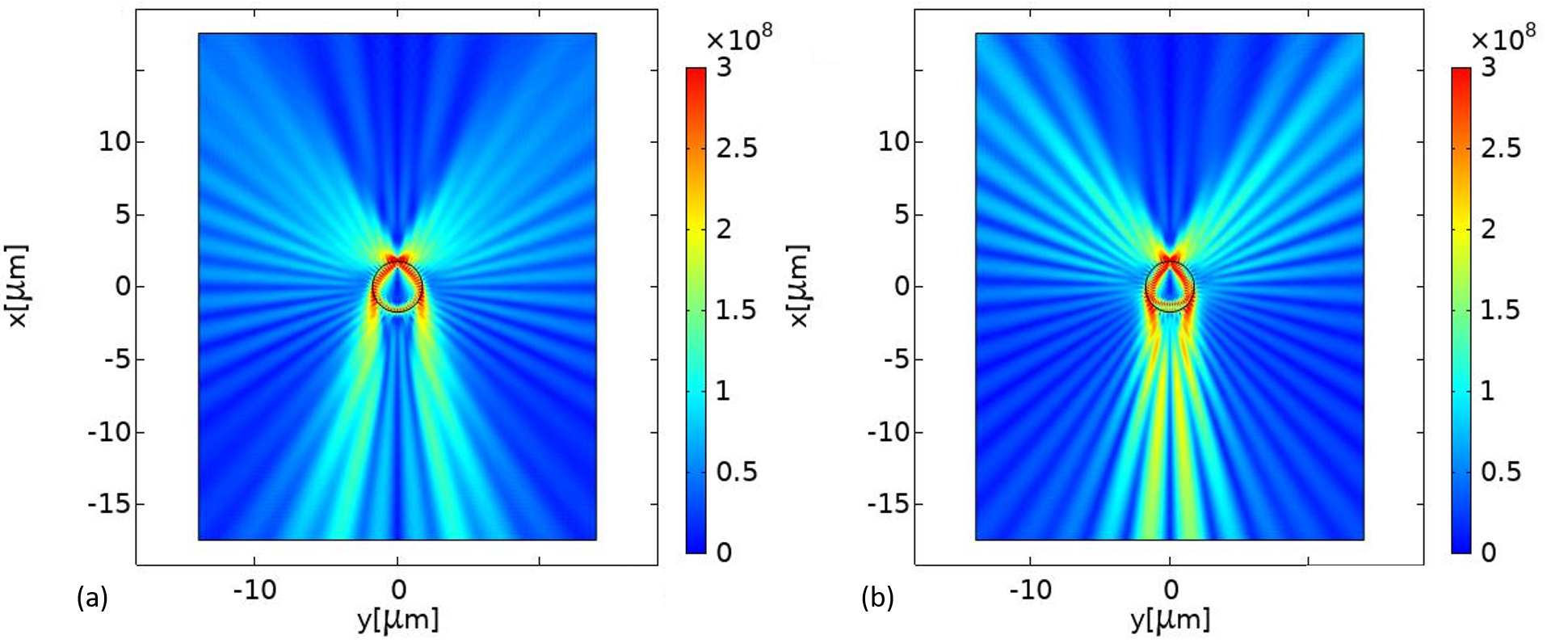}
    \caption{Single source -- intensity maps in the large area for $n=1.4$ (a) and $n=1.7$ (b). Here $kR=20$ and $\lambda=550$ nm.}
    \label{Pic8}
    \end{figure*}

In Fig.~\ref{Pic8} we present a comparison of the large-area intensity maps for the cases $kR=20$, $n=1.4$ and $kR=20$, $n=1.7$. In the first case, the imaging beam mimics the function $\partial J_{\nu}(15\theta)/\partial \nu$ at $\nu=1.5$ and in the second case -- the function $J_1(\xi\theta)$, where $\xi=31$. In both cases, there are no features of diffraction.

\subsection{Simulations for a dual normally polarized source}

\begin{figure*}[h!]
\centering
\includegraphics[width=0.95\textwidth]{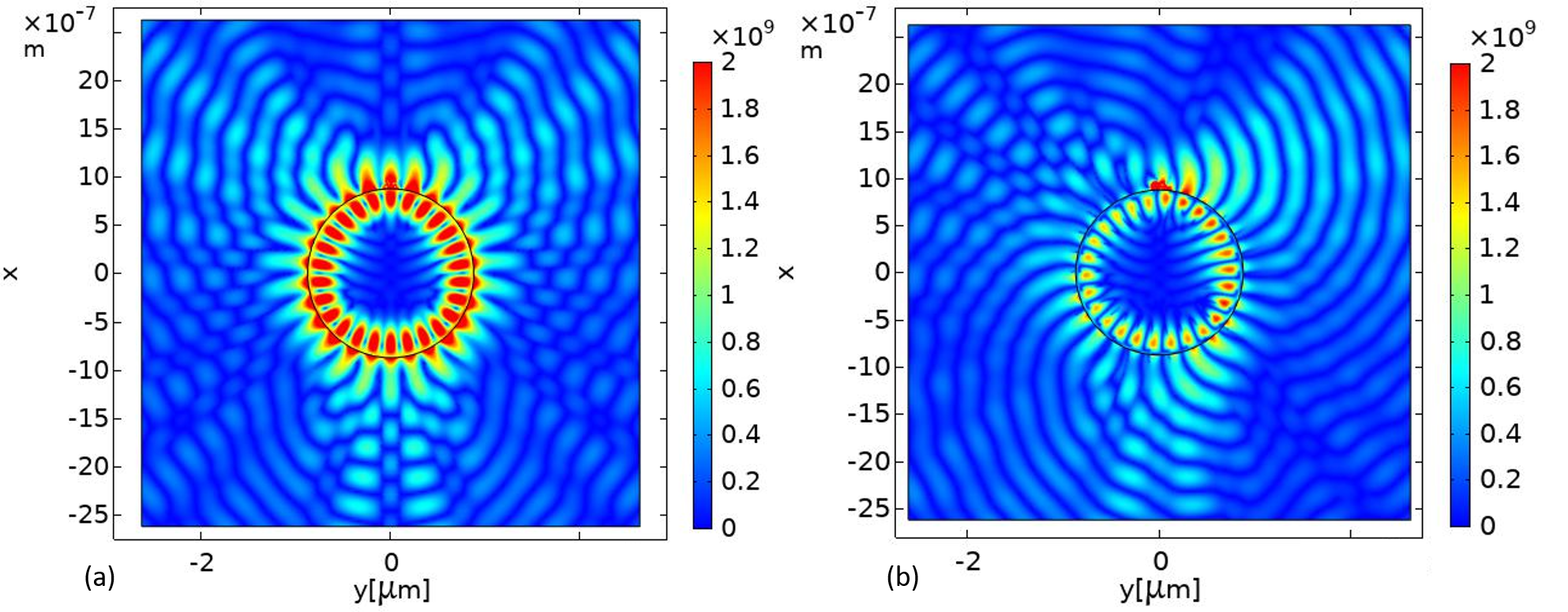}
\caption{Dual source -- wave pictures for the symmetric ($\phi=0$) case (a) and asymmetric ($\phi=k\delta$) case  case (b).
Here $kR=10$ and $n=1.7$.}
\label{Pic9}
\end{figure*}

We have performed similar simulations for a symmetric (in-phase) dual source with $\delta=2d=1/k$ and observed no changes but a slight angular extension of the partial beams by the angle close to $\delta/R$. Below we concentrate on the case of the asymmetric dual source $p_2=p_1\exp(ik\delta)$.
The impact of the phase asymmetry is especially spectacular when $n$ is larger and $kR$ is smaller. In Fig.~\ref{Pic9} we compare the wave pictures obtained for a symmetric dual source and an asymmetric one when $n=1.7$ and $kR=10$. Since in this case $\Psi>\pi$  we observe a
standing wave pattern in the microparticle that resembles a whispering gallery resonance (though it is not so). in the symmetric case (as well as for a single source) it results in strongly dominating partial beams of the first order. In the asymmetric case, this picture is drastically modified.
Even from the wave picture in the Fresnel zone it is clear that the image of the asymmetric source should be qualitatively different from that of a symmetric one.

\begin{figure*}[h!]
\centering
\includegraphics[width=0.95\textwidth]{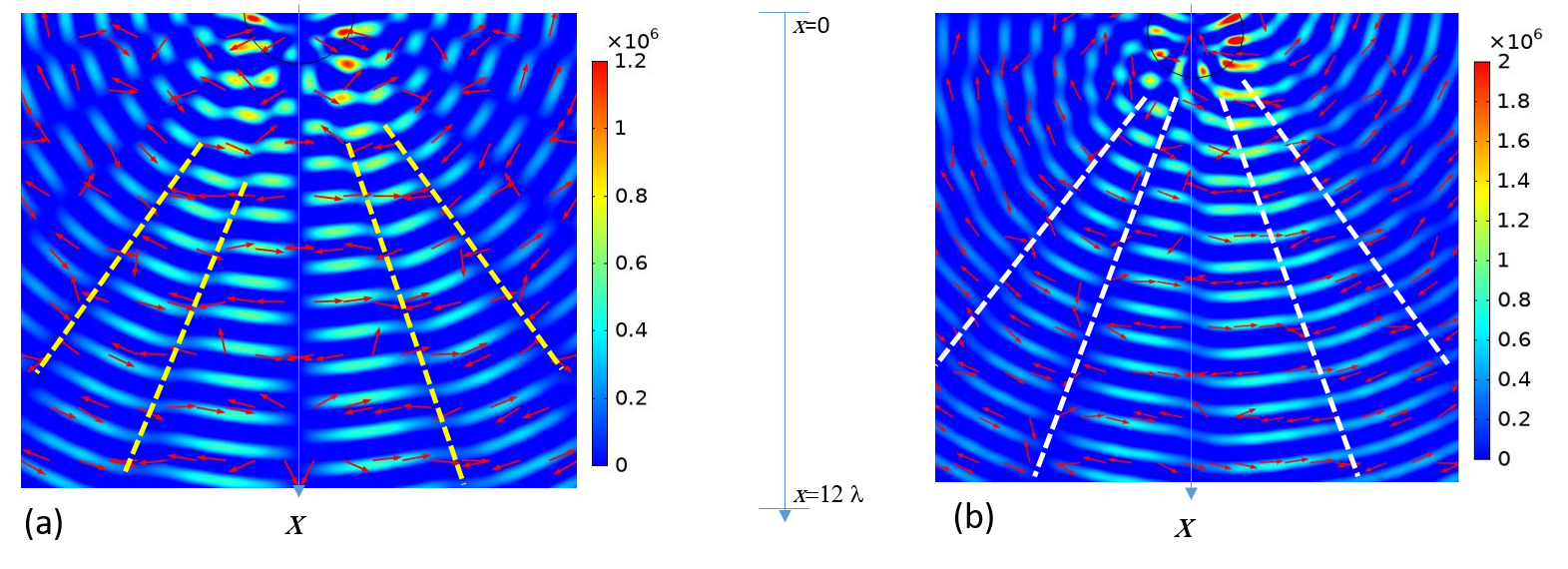}
\caption{Single source (a) and asymmetric dual (b) source -- vector distributions and wave pictures for the case $kR=10$, $n=1.4$.
Dashed lines show the bounds of partial beams. The color map represents the instantaneous magnetic field and arrows show the direction of total electric field. The color map for a dual source is brighter because the total dipole moment is larger.
}
\label{Pic10}
\end{figure*}

In Fig.~\ref{Pic10} we compare the spatial distributions of the vector \textbf{E} (shown by arrows on the background of a wave picture) for two cases: (a) a single dipole source and (b) a dual asymmetric source. In these pictures, the bounds of partial beams $m=\pm 1,\pm  2$ are shown by dashed lines. For the dual source we see that the angular width of partial beams is slightly extended (nearly by $\delta/R$, as predicted by the theory). The phase of the electric field on the rays propagating with the same tilt $\theta$ to the $x$-axis in the left and right halves of the plot depicted Fig.~\ref{Pic10}(b) are clearly different. Visually it is impossible to estimate this difference, but qualitatively, these observations confirm the theoretical expectations. Note, that in Fig.~\ref{Pic10} the color map shows only positive maxima of H (those where $H_z>0$). Therefore the distance between the bright areas of picture is here equal $\lambda$ (not $\lambda/2$ as in the wave pictures above). This is also the reason why the bright areas in the left ($y<0$) and right($y>0$) halves are shifted by $\lambda/2$.

\begin{figure*}[h!]
\centering
\includegraphics[width=0.95\textwidth]{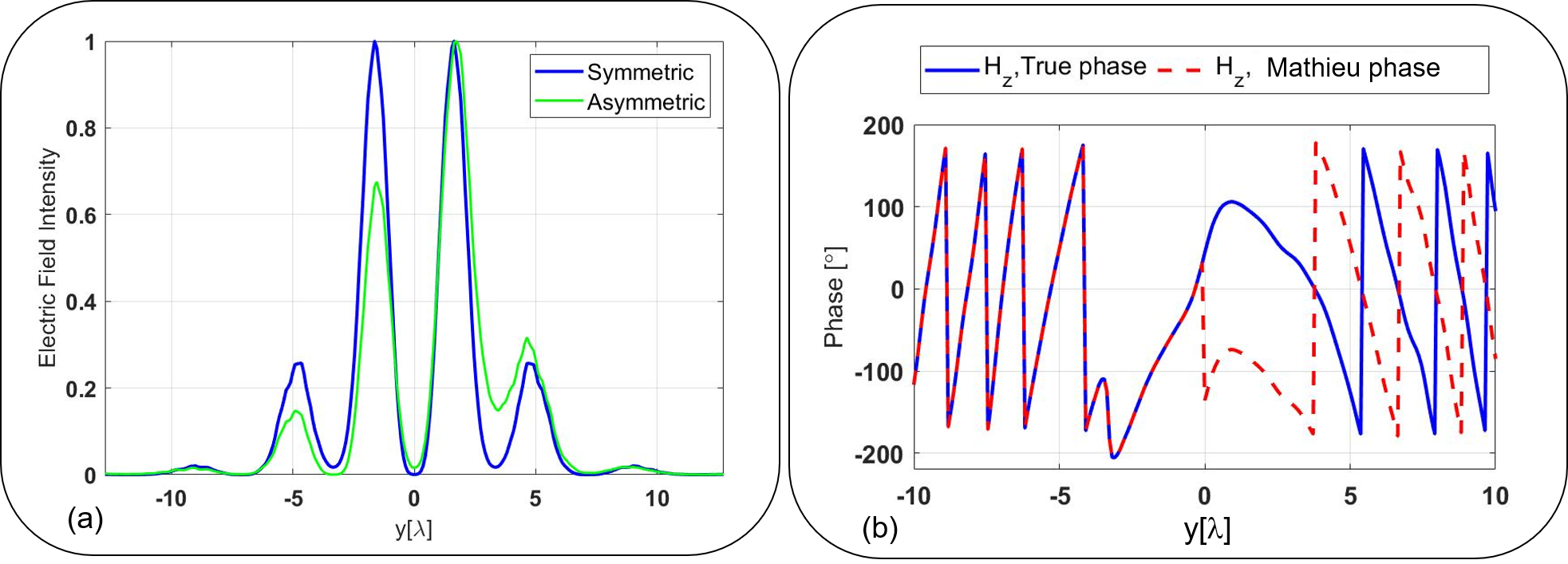}
\caption{Asymmetric dual source. Intensity (a) and phase (b) distributions across six partial beams for the case $kR=10$, $n=1.4$.}
\label{Pic11}
\end{figure*}

Finally, in Fig.~\ref{Pic11} we present typical distributions of intensity and phase across an imaging beam. The distributions corresponds to the case $n=1.4$, $kR=10$ (Mathieu beam) and in this example the axis $y$ crosses the partial beams $m=\pm 1,\pm 2,\pm 3$ on the same distance as in Fig.~\ref{Pic6}(a). For comparison, in Fig.~\ref{Pic11}(a) we show also the intensity distribution corresponding to the symmetric dual source. This distribution corresponds to the concept of the Mathieu beam and the asymmetric source grants the interference of two Mathieu beams. Comparing Figs.~\ref{Pic11}(b) with the symmetric counterpart depicted in
Figs.~\ref{Pic6}(b) we see that the theoretic expectations are fully confirmed by simulations. The even function of $y$ is replaced by an asymmetric one, and the Matheiu phase taken at the point $y$ differs from that taken at $(-y)$ by nearly $\phi$. Similar calculations were performed for $kR=20-30$ ($n=1.4$) and for $kR=10-30$ ($n=1.7$) and we saw that this phase asymmetry holds in all cases for Mathieu imaging beams, Bessel beams and a newly-revealed diffraction-free beam which emulates the Bessel function derivative over the index. We have also checked that the extension of the phase shift from $\phi=k\delta$ to $\phi=k\delta(n+1)/2$ keeps the mechanism illustrated by Fig.~\ref{Pic4}(b). In this case, the Mathieu or Bessel phase difference for two equally tilted rays corresponding to $+m$ and $-m$ increases from $\phi=k\delta$ to $\phi=k\delta(n+1)/2$.
This corresponds to the increase of the distance $2\Delta y$ between points $I_1$ and $I_2$. On the contrary, the reduction of $\phi$ decreases the phase shift between two symmetrically tilted rays (and should, therefore, decrease the magnification).

\subsection{Discussion}

In this work, we have concentrated on the superlens operation in the regime when CWs are efficiently excited by a dipole source polarized normally with respect to the MP surface. Since the spectrum of CWs is discrete and finite they eject from the MP surface forming a symmetric set of partial beams that we treat as an imaging beam. The polarization of this beam is anti-symmetric with respect to the axis drawn between the source and the particle center. The distribution of the phase and amplitude of the field across the imaging beam allowed us to assume that the imaging beam is practically diffraction-free. This assumption was confirmed by exact simulations. Depending on the MP radius the imaging beam turned out to be either the Bessel beam or the Mathieu beam.

Unfortunately, the absence of diffraction does not guarantee the subwavelength resolution. Two point-wise scatterers with the subwavelength gap $\delta\ll\lambda$ between them create in the regime of CWs an imaging beam that has practically the same intensity distribution as that created by one (total) dipole located in the middle of the gap. They cannot be resolved if they are in-phase. However, if this dual source is coherent and comprises a phase shift proportional to $\delta$, in a plane shifted forward with respect to the image plane of the total dipole  two maxima of intensity arise at two sides of the optical axis. The gap between these maxima is proportional to $\delta$.

The key hypotheses referring to the CW regime were confirmed by the numerical simulations. Here we do not report only the image resulting from the CWs. Also, we do not present a study of the finest possible resolution in this regime. These issues will be reported in our next papers.
To simulate the structures we have simulated above adding a 2D lens in a COMSOL project is possible. However, in order to obtain the subwavelength image we should suppress besides of the diffraction also the aberrations. This task is difficult and it is not reasonable to add it to the goals of the present paper.

Instead, let us analyze the literature data having in mind the mechanism of the subwavelength imaging suggested above. In work \cite{NC} the superlens operation of a barium titanate microsphere in the case of a coherent illumination by a laser light was experimentally studied. Here the scatterers to be resolved represented the grooves and notches in the silicon substrate on which the MP was located. Two cases of the wave incidence were studied: a symmetric one (along the $x$ axis in our notations) and an asymmetric one ($20^{\circ}$ to the $y$ axis). For the symmetric excitation the subwavelength ($\delta<\lambda/2$)
resolution of two scatterers was obtained for scatterers located in a circle of radius $0.8\, \mu$m  centered by the point where the sphere touched the patterned substrate.
Since the sphere in this experiment had the radius $R=27\, \mu$m all points of this circle are distanced by $10$ nm or less from the surface of the MP.
This subwavelength imaging can hardly be treated as that granted by a sphere. The mechanism of this imaging is governed by the properties of a tiny air crevice between two
highly refractive materials.

The asymmetric incidence offers the enlargement of the circle in which the subwavelength resolution is observed. In the asymmetric case this radius is twofold and the distance between the sphere and the substrate increases from $10$ nm up to $40$ nm. In the outer part of the circle the field concentration is not so high, and the spherical profile of the MP becomes important. For the sources located in this area the mechanism assumed in the present paper can prevail. Sources located further from the touching point are not resolved simply because they are too distant from the sphere and therefore do not excite the CWs so efficiently.

\section{Conclusions}

In the present work, we suggested and discussed some possible non-resonant mechanisms of the superlens operation of a dielectric microscphere or microcylinder. We claim that there are several mechanisms of nanoimaging granted by these MPs and that, contrarily to the popular opinion, some of them (if not all) have nothing to do with the phenomenon of a photonic nanojet.

In the present paper, we consider the MP without a substrate. Then, to our opinion, a non-resonant superlens operation of a MP is granted by the normal polarization of the object with respect to the MP. There are two mechanisms of the far-field subwavelength imaging -- a coherent one and an incoherent one.
A coherent mechanism (when the image results from the set of creeping waves) is considered here in more details. A non-coherent one (when the image is created by the radially polarized beam) is concerned briefly. In our next papers, we plan to continue the study of the coherent mechanism and to prove the feasibility of the incoherent one.

\section*{ACKNOWLEDGMENTS}

~~~~~~This work was funded through the EMPIR project 17FUN01-BeCOMe. The EMPIR initiative is co-funded by
the European Union Horizon 2020 research and innovation programme and the EMPIR participating States.


\section*{References}


\begin{thebibliography}{00}




\bibitem{STED}
Rittweger E, Han K Y, Irvine S E, Eggeling C and Hell S W 2009 microscopy reveals crystal colour centres with nanometric resolution {\it Nat. Photon.} {\bf 3} 144–7


\bibitem{So}
So S, Kim M, Lee D, Nguyen D M and Rho J 2018 Overcoming diffraction limit: From microscopy to
nanoscopy {\it Appl. Spectrosc. Rev.} {\bf 53} 290-312

\bibitem{Hong}
Wang Z, Guo W, Li L, Lukyanchuk B, Khan A, Liu Z, Chen Z and Hong M 2011
Optical virtual imaging at 50 nm lateral resolution with a white-light nanoscope {\it Nat. Commun.} {\bf 2} 218

\bibitem{PRB}
Mollaei M S M and Simovski C R 2019 Dual-metasurface superlens: A comprehensive study {\it Phys. Rev. B} {\bf 100} 205426

\bibitem{PNJ}
Luk’yanchuk B S, Wang Z B, Song W D and Hong M H 2004 Particle on surface: 3D-effects in dry laser cleaning {\it Appl. Phys. A} {\bf 79} 747–51

\bibitem{Taflove}
Chen Z, Taflove A and Backman V 2004 Photonic nanojet enhancement of backscattering of light by nanoparticles: a potential novel visible-light ultramicroscopy technique
{\it Opt. Express} {\bf 12} 1214–20

\bibitem{Itagi}
Itagi A V and Challener W A 2005 Optics of photonic nanojets {\it J. Opt. Soc. America A} {\bf 22} 2847-58

\bibitem{Taflove1}
Heifetz A, Kong S C, Sahakian A V, Taflove A and Backman V 2009 Photonic nanojets {\it J. Comput. Theor. Nanosci.} {\bf 6} 1979–92

\bibitem{Yang}
Yang H, Trouillon R, Huszka G and Gijs M A M 2016 Super-resolution imaging of a dielectric microsphere is governed by the waist of its photonic nanojet {\it NanoLetters} {\bf 16} 4862-66

\bibitem{Lecler}
Lecler S, Perrin S, Leong-Hoi A. and Montgomery P 2019 Photonic jet lens {\it Scientific Reports} {\bf 9} 4725

\bibitem{Hyper1}
Liu Z, Lee H, Xiong Y, Sun C and Zhang X 2007 Far-field optical hyperlens magnifying sub-diffraction-limited
objects {\it Science} {\bf 315} 1686

\bibitem{Hyper2}
Zhang X and Liu Z 2008 Superlenses to overcome the diffraction limit {\it Nature Mat.} {\bf 7} 425-30

\bibitem{Hyper3}
Rho J, Ye Z, Xiong Y, Yin X, Liu Z, Choi H, Bartal G and Zhang X 2010 Spherical hyperlens for two-dimensional sub-diffractional imaging at visible frequencies {\it Nature Comm.} {\bf 1} 143

\bibitem{Hyper4}
Lu D and Liu Z 2012 Hyperlenses and metalenses for far-field super-resolution imaging {\it Nature. Comm.} {\bf 3} 1205

\bibitem{Kassamakov}
Kassamakov I, Lecler S, Nolvi A, Leong-Hoi A, Montgomery P and Haeggstr{\"o}m E 2016
3D super-resolution optical profiling using microsphere-enhanced Mirau interferometry {\it Sci. Rep.} {\bf 7} 3683

\bibitem{Astratov}
Allen K W, Farahi N, Li Y, Limberopoulos N I, Walker D E, Urbas A M and Astratov V N 2015
Overcoming the diffraction limit of imaging nanoplasmonic arrays by microspheres and microfibers  {\it Opt. Express} {\bf 23} 24484

\bibitem{Astratov1}
Maslov A V and Astratov V N 2019 Resolution and reciprocity in microspherical nanoscopy: point-spread function versus photonic nanojets
{\it Phys. Rev. Applied} {\bf 11} 064004

\bibitem{Zhou}
Zhou S, Deng Y, Zhou W, Yu M, Urbach H P and Wu Y 2017 Effects of whispering gallery mode in microsphere super-resolution imaging {\it Appl. Phys. B} {\bf 123} 236

\bibitem{Maslov}
Maslov A V and Astratov V N 2017 Optical nanoscopy with contact Mie-particles: Resolution analysis  {\it Appl. Phys. Lett.} {\bf 110} 261107

\bibitem{Cang}
Cang H, Salandrino A, Wang Y and Zhang X 2015 Adiabatic far-field sub-diffraction imaging
{\it Nature Comm.} {\bf 6} 7942

\bibitem{Wang}
Wang Z, Luk’yanchuk B, Yue L, Paniagua-Domínguez R, Yan B, Monks J, Minin O V, Minin I V, Huang S and Fedyanin A A 2019 Super-resonances in microspheres: extreme effects in field localization  arXiv:1906.09636 [phys. Opt.]


\bibitem{Sundaram}
Sundaram V M and Wen S-B Analysis of deep sub-micron resolution in microsphere based imaging 2014 {\it Appl. Phys. Lett.} {\bf 105} 204102

\bibitem{Tidwell}
Tidwell S C, Ford D H, and Kimura W D, Generating radially polarized beams interferometrically, 1990 {\it Appl. Optics} {\bf 29} 2234-2239

\bibitem{radial}
Zhu Q X 2009 Description of the propagation of a radially polarized beam with the scalar Kirchhoff diffraction {\it J. Mod. Optics} {\bf 56} 1621–25

\bibitem{Felsen}
Heyman E and Felsen L B 1984 High-frequency fields in the presence of a curved dielectric interface {\it IEEE Trans. Antennas Propag.} {\bf 32} 969-78

\bibitem{Born}
Born M and Wolf E 2003 {\it Principles of Optics} 7th Ed (UK: Cambridge University Press)

\bibitem{Papaliolios}
Rueckner W and Papaliolios C 2001 How to beat the Rayleigh resolution limit: A lecture demonstration {\it Am. J. Phys.} {\bf 70} 587-94


\bibitem{Osipov}
Osipov A V and Tretyakov S A 2017 {\it Modern Electromagnetic Scattering Theory with Applications} (Wiley, Chichetser, UK) pp 446-47, 454-56, 535-41

\bibitem{Narimanov}
Narimaov E 2019 Resolution limit of label-free far-field microscopy {\it Adv. Photon.} {\bf 1} 056003

\bibitem{Simon}
Simon D S 2016 {\it A Guided Tour of Light Beams} (Bristol UK: IOP Publishing)


\bibitem{NC}
Wang F, Liu L, Yu H, Wen Y, Yu P, Liu Z, Wang Y and Li W J 2016 Scanning superlens microscopy for non-invasive large field-of-view visible light nanoscale imaging {\it Nature Comm.} {\bf 7}  13748











\end{thebibliography}
\end{document}